\documentclass{article}

\usepackage{fullpage}
\usepackage{amsmath}
\usepackage{bm}

\let\oldbibliography\thebibliography
\renewcommand{\thebibliography}[1]{%
  \oldbibliography{#1}%
  \setlength{\itemsep}{0pt}%
}

\newcommand{\ket}[1]{\vert #1 \rangle}
\newcommand{\proj}[1]{\ket{#1}\langle #1 \vert}
\newcommand{\Tr}[1]{\text{Tr} \left ( #1 \right )}

\title{``It from bit'' and the quantum probability rule}

\author{M. S. Leifer}

\date{May 17, 2014}

\begin{document}

\maketitle

\begin{abstract}
  I argue that, on the subjective Bayesian interpretation of
  probability, ``it from bit'' requires a generalization of
  probability theory.  This does not get us all the way to the quantum
  probability rule because an extra constraint, known as
  noncontextuality, is required.  I outline the prospects for a
  derivation of noncontextuality within this approach and argue that
  it requires a realist approach to physics, or ``bit from it''.  I
  then explain why this does not conflict with ``it from bit''.  This
  version of the essay includes an addendum responding to the open
  discussion that occurred on the FQXi website.  It is otherwise
  identical to the version submitted to the contest.
\end{abstract}

\section{Wheeler's ``it from bit''}

\begin{quotation}
  It from bit. Otherwise put, every it---every particle, every field
  of force, even the spacetime continuum itself---derives its
  function, its meaning, its very existence entirely---even if in some
  contexts indirectly---from the apparatus-elicited answers to yes or
  no questions, binary choices, bits.

  It from bit symbolizes the idea that every item of the physical
  world has at bottom---at a very deep bottom, in most instances---an
  immaterial source and explanation; that what we call reality arises
  in the last analysis from the posing of yes-no questions and the
  registering of equipment-evoked responses; in short, that all things
  physical are information-theoretic in origin and this is a
  participatory universe.

  --- J. A. Wheeler \cite{Wheeler1990}
\end{quotation}

John Wheeler's ``it from bit'' is a thesis about the foundations of
quantum theory.  It says that the things that we usually think of as
real---particles, fields and even spacetime---have no existence
independent of the questions that we ask about them.  When a detector
clicks it is not registering something that was there independently of
the experiment.  Rather, the very act of setting up the detector in a
certain way---the choice of question---is responsible for the
occurrence of the click.  It is only the act of asking questions that
causes the answers to come into being.  This idea is perhaps best
illustrated by Wheeler's parable of the game of twenty questions
(surprise version).

\begin{quotation}
  You recall how it goes---one of the after-dinner party sent out of
  the living room, the others agreeing on a word, the one fated to be
  a questioner returning and starting his questions. ``Is it a living
  object?'' ``No.'' ``Is it here on earth?'' ``Yes.'' So the questions
  go from respondent to respondent around the room until at length the
  word emerges: victory if in twenty tries or less; otherwise, defeat.

  Then comes the moment when we are fourth to be sent from the
  room. We are locked out unbelievably long. On finally being
  readmitted, we find a smile on everyone's face, sign of a joke or a
  plot. We innocently start our questions. At first the answers come
  quickly. Then each question begins to take longer in the
  answering---strange, when the answer itself is only a simple ``yes''
  or ``no.'' At length, feeling hot on the trail, we ask, ``Is the
  word `cloud' ?'' ``Yes,'' comes the reply, and everyone bursts out
  laughing.  When we were out of the room, they explain, they had
  agreed not to agree in advance on any word at all. Each one around
  the circle could respond ``yes'' or ``no'' as he pleased to whatever
  question we put to him. But however he replied he had to have a word
  in mind compatible with his own reply---and with all the replies
  that went before. No wonder some of those decisions between ``yes''
  and ``no'' proved so hard!

  --- J. A. Wheeler \cite{Wheeler1979}
\end{quotation}

Wheeler proposed ``it from bit'' as a clue to help us answer the
question ``How come the quantum?'', i.e. to derive the mathematical
apparatus of quantum theory from a set of clear physical principles.
In this essay, I discuss whether ``it from bit'' implies the quantum
probability rule, otherwise known as the Born rule, which would get us
part of the way towards answering Wheeler's question.

My main argument is that, on the subjective Bayesian interpretation of
probability, ``it from bit'' requires a generalized probability
theory.  I explain why this is not ruled out by the common claim that
classical probability theory is not to be violated on pain of
irrationality.

In the context of quantum theory, ``it from bit'' does not quite get
us all the way to the Born rule because the latter mandates a further
constraint known as noncontextuality.  The prospects for understanding
noncontextuality as a rationality requirement or an empirical addition
are slim.  Extra physical principles are needed and I argue that these
must be about the nature of reality, rather than the nature of our
knowledge.  This seems to conflict with ``it from bit'' as it requires
and agent-independent reality, suggesting ``bit from it''.  I argue
that there is no such conflict because the sense of ``it'' used in
``it from bit'' is different from the sense used in ``bit from it''.

\section{The interpretation of probability}

Since von Neumann's work on quantum logic and operator algebras
\cite{Birkhoff1936, Murray1936}, it has been known that quantum theory
can be viewed as a generalization of probability theory
\cite{Redei2007, Leifer2011}.  If we want to understand what this
tells us about the nature of reality then we will need to adopt a
concrete theory of how probabilities relate to the world, which is the
job of an interpretation of probability theory\footnote{See
  \cite{Gillies2000} for an accessible introduction and
  \cite{Eagle2011} for a collection of key papers.}.  Three main
classes of interpretation have arisen to meet this need: frequentism
(probability is long run relative frequency), epistemic probability
(probabilities represent the knowledge, information, or beliefs of a
decision making agent), and objective chance (probabilities represent
a kind of law of nature or a disposition for a system to act in a
certain way).  Getting into the details of all these options would
take us too far afield, but a few comments are in order to explain why
adopting my preferred epistemic interpretation, known as subjective
Bayesianism\footnote{Subjective Bayesianism has its origins in
  \cite{Finetti1931, Ramsey1931}.  An accessible introduction is
  \cite{Jeffrey2004}.}, is not a crazy thing to do.

Frequentism is still popular amongst physicists, but it has largely
been abandoned by scholars of the philosophy of probability.  It is
not able to handle single-case probabilities, e.g. the probability
that civilization will be destroyed by a nuclear war, and it leads to
a bizarre reading of the law of large numbers that does not do the
explanatory work required of it\footnote{See \cite{Howson2005} for a
  critique of frequentism in statistics.}.  A common position in the
philosophy of probability is that subjective Bayesianism is more
satisfactory, but that it needs to be backed up by some theory of
objective chance in order to account for probabilistic
laws\footnote{This view originates with David Lewis
  \cite{Lewis1980}.}.  My own view is that subjective Bayesianism
suffices on its own, but whether or not one believes in objective
chance is irrelevant for the present discussion, since objective
chances need to be connected to epistemic probabilities in some way in
order to explain how we can come to know statistical laws.  The usual
way of doing this is via David Lewis' principal principle
\cite{Lewis1980}.  One of the implications of this is that objective
chances must have the same mathematical structure as subjective
Bayesian probabilities.  Therefore, if we can argue that ``it from
bit'' requires a modification of subjective Bayesian probability then
the same will apply to objective chances as well.  It is also worth
noting that several modern interpretations of quantum theory adopt
subjective Bayesianism, including ``Quantum Bayesianism''
\cite{Caves2002, Fuchs2003, Fuchs2010, Fuchs2010a} and the
Deutsch-Wallace variant of many-worlds \cite{Deutsch1999, Wallace2007,
  Wallace2010, Wallace2012} amongst others \cite{Pitowsky2003}.

\section{Subjective Bayesian probability}

Subjective Bayesianism says that probabilities represent the degrees
of belief of a decision making agent, who is conventionally described
in the second person as ``you''.  Degrees of belief are measured by
looking at your behaviour, e.g. your willingness to enter into bets.
The claim is that if you do not structure your beliefs according to
the axioms of probability theory then you are irrational.  There are
various ways of deriving this, differing in their simplicity and
sophistication.  For ease of exposition, I base my discussion on the
simplest approach, known as the Dutch book argument.

The Dutch book argument defines your degree of belief in the
occurrence of an uncertain event $E$ as the value $\$p(E)$ you
consider to be a fair price for a lottery ticket that pays $\$1$ if
$E$ occurs and nothing if it does not.  ``Fair price'' here means that
you would be prepared to buy or sell any number of these tickets at
that price and that you would be prepared to do this in combination
with fair bets on arbitrary sets of other events.  Your degrees of
belief are said to be irrational if a malicious bookmaker can force
you to enter into a system of bets that would cause you to lose money
whatever the outcome, despite the fact that you consider them all
fair.  Otherwise, your degrees of belief are said to be rational.  The
Dutch book argument then shows that your degrees of belief are
rational if, and only if, they satisfy the usual axioms of probability
theory.  These axioms are:

\begin{itemize}
\item \textbf{Background framework}: There is a set $\Omega$, called
  the sample space, containing the most fine-grained events you might
  be asked to bet on, e.g. if you are betting on the outcome of a dice
  roll then $\Omega = \{1,2,3,4,5,6\}$. In general, an event is a
  subset of $\Omega$, e.g. the event that the dice roll comes out odd
  is $\{1,3,5\}$.  For simplicity, we assume that $\Omega$ is finite.
  The set of events forms a Boolean algebra, which just means that it
  supports the usual logical notions of AND, OR and NOT.
\item \textbf{A1}: For all events $E \subseteq \Omega$, $0 \leq p(E)
  \leq 1$.
\item \textbf{A2}: For the certain event $\Omega$, $p(\Omega) = 1$.
\item \textbf{A3}: If $E \cap F = \emptyset$, i.e. $E$ and $F$ cannot
  both happen together, then $p(E \cup F) = p(E) + p(F)$, where $E
  \cup F$ means the event that either $E$ or $F$ occurs.
\end{itemize}

For illustrative purposes, here is the part of the argument showing
that violations of \textbf{A1} and \textbf{A2} are irrational.
Consider an event $E$ and suppose contra \textbf{A1} that $p(E) < 0$.
This means that you would be willing to sell a lottery ticket that
pays out on $E$ to the bookie for a negative amount of money, i.e. you
would pay her $\$p(E)$ to take the ticket off your hands.  Now, if $E$
occurs you will have to pay the bookie $\$1$ so in total you will have
paid her $\$1+p(E)$, and if $E$ does not occur you will have paid her
a total of $\$p(E)$.  Either way, you will lose money so having
negative degrees of belief is irrational.  A similar argument shows
that having degrees of belief larger than $1$ is irrational.  Now
suppose contra \textbf{A2} that $p(\Omega) < 1$.  Then, you would be
prepared to sell the lottery ticket for $\$p(\Omega)$ and pay out
$\$1$ if $\Omega$ occurs.  However, since $\Omega$ is certain to
occur, you will always end up paying out, which leaves you with a
loss.

\section{Is probability theory normative?}

Based on this kind of argument, many subjective Bayesians regard
probability theory as akin to propositional logic\footnote{For
  example, see \cite{Finetti2008} where this argument is made
  repeatedly.}.  In logic, you start with a set of premises that you
regard as true, and then you use the rules of logic to figure out what
other propositions must be true or false as a consequence.  If you
fail to abide by those truth values then there is an inconsistency in
your reasoning.  However, there is nothing in logic that tells you
what premises you have to start with.  The premises are simply the
input to the problem and logic tells you what else they compel you to
believe.  Similarly, subjective probability does not tell you what
numerical value you must assign to any uncertain event\footnote{This
  is where it differs from objective Bayesianism \cite{Jaynes2003},
  which asserts that there is a unique rational probability that you
  ought to assign.  However, defining such a unique probability is
  problematic at best.}, but given some of those values as input, it
tells you what values you must assign to other events on pain of
inconsistency, the inconsistency here being exposed in the form of a
sure loss.  Like logic, subjective Bayesians regard probability theory
as normative rather than descriptive, i.e. they claim that you should
structure your degrees of belief about uncertain events according to
probability theory if you aspire to be ideally rational, but not that
humans actually do structure their beliefs in this way.  In fact, much
research shows that they do not \cite{Kahneman2012}.

The normative view of probability theory presents a problem if we want
to view quantum theory as generalized probability because it implies
that it is irrational to use anything other than conventional
probability theory to reason about uncertain events.  Fortunately, the
normative view is not just wrong, but obviously wrong.  Unlike logic,
it is easy to come up with situations in which the Dutch book argument
has no normative force.  Because of this, the idea that it might
happen in quantum theory too is not particularly radical.

For example, the Dutch book argument requires that you view the fair
price for selling a lottery ticket to be the same as the fair price
for buying it.  In reality, people are more reluctant to take on risk
than they are to maintain a risk for which they have already paid the
cost.  Therefore, the fair selling price might be higher than the fair
buying price.  This leads to the more general theory of upper and
lower probabilities wherein degrees of belief are represented by
intervals on the real line rather than precise numerical values
\cite{Smith1961}.

At this point, I should address the fact that the Dutch book argument
is not the only subjective Bayesian derivation of probability theory,
so its defects may not be shared with the other derivations.  The most
general subjective arguments for probability theory are formulated in
the context of decision theory, with Savage's axioms being the most
prominent example \cite{Savage1972}.  These take account of things
like the fact that you may be risk averse and your appreciation of
money is nonlinear, e.g. $\$1$ is worth more to a homeless person than
a billionaire, so they replace the financial considerations of the
Dutch book argument with the more general concept of ``utility''.
However, what all these arguments have in common is that they are
hedging strategies.  They start with some set of uncertain events and
then they introduce various decision scenarios that you could be faced
with where the consequences depend on uncertain events in some way,
e.g. the prizes in a game that depends on dice rolls.  Importantly,
these arguments only work if the set of decision scenarios is rich
enough.  They ask you to consider situations in which the prizes for
the various outcomes are chopped up and exchanged with each other in
various ways.  For example, in the Dutch book argument this comes in
the form of the idea that you must be prepared to buy or sell
arbitrarily many tickets for arbitrary sets of events at the fair
price.  The arguments then conclude that if you do not structure your
beliefs according to probability theory then there is some decision
scenario in which you would be better off had you done so and none in
which you would be worse off.  However, in real life, it is rather
implausible that you would be faced with such a rich set of decision
scenarios.  More often, you know something in advance about what
decisions you are going to be faced with.  This is why decision
theoretic arguments are hedging strategies.  They start from a
situation in which you do not know what decisions you are going to be
faced with and then they ask you to consider the worst possible
scenario.  If you know for sure that this scenario is not going to
come up then the arguments have no normative force.

As an example, consider the following scenario.  There is a coin that
is going to be flipped exactly once.  You have in your possession
$\$1$ and you are going to be forced to bet that dollar on whether the
coin will come up heads or tails, with a prize of $\$2$ if you get it
right.  You do not have the option of not placing a bet.  How should
you structure your beliefs about whether the coin will come up heads
or tails?  If the decision theoretic arguments applied then we would
be forced to say that you must come up with a precise numerical value
for the probability of heads $p(H)$.  However, it is clear that the
cogitation involved in coming up with this number is completely
pointless in this scenario.  All you need to know is the answer to a
single question.  Do you think heads is more likely to come up than
tails?  Your decision is completely determined by this answer, which
is just a single bit of information rather than a precise numerical
value.

It should be clear from this that the decision theoretic arguments are
not as strongly normative as the laws of logic.  Instead they are
\emph{conditionally} normative, i.e. normative if the decision
scenarios envisaged in the argument are all possible.

\section{``It from bit'' implies a generalized probability theory}

A universe that obeys ``it from bit'' is a universe in which not all
conceivable decision scenarios are possible.  To explain this,
consider again Wheeler's parable of twenty questions (surprise
version) and imagine that you are observing the game passively,
placing bets with a bookmaker on the side as it proceeds.  To make
things more analogous to quantum theory, imagine that the respondents
exit the room as soon as they have answered their question, never to
be heard from again.  We might imagine that they are sent through a
wormhole into a region of spacetime that will forever be outside of
our observable universe and that the wormhole promptly closes as soon
as they enter it.  This rules out the possibility that we might ask
them about what they would have answered if they had been asked a
different question, since in quantum theory we generally cannot find
out what the outcome of a measurement that we did not actually make
would have been.

Suppose that, at some point in the game, you make a bet with the
bookie that the object that the fifth respondent has in mind is a
dove.  However, what actually happens is that the questioner asks ``Is
it white?'' and the answer comes back ``yes'', whereupon the fifth
respondent is whisked off to the far corners of the universe.  Now,
although the answer ``yes'' is consistent with the object being a
dove, this is not enough to resolve the bet as there are plenty of
other conceivable white objects.  As in Wheeler's story, suppose that
the last question asked is ``Is it a cloud?'' and that the answer
comes back ``yes''.  In the usual version of twenty questions this
would be enough to resolve the bet in the bookie's favor because all
the respondents are thinking of a common object.  However, in the
surprise version this is not the case.  It could well be that ``dove''
was consistent with all the answers given so far at the time we made
the bet, and that the fifth respondent was actually thinking of a
dove.  We can never know and so the bet can never be resolved.  It has
to be called off and you should get a refund.

Whilst the bet described above is unresolvable, other bets are still
jointly resolvable, e.g. a bet on whether the fifth respondent was
thinking of a white object together with a bet on whether the last
respondent was thinking of a cloud.  The set of bets that is jointly
resolvable depends on the sequence of questions that is actually asked
by the questioner.  If you want to develop a hedging strategy ahead of
time, then you need to consider all possible sequences of questions
that might be asked to ensure that you cannot be forced into a sure
loss for any of them.

For the subjective Bayesian, the main lesson of this is that, in
general, only certain subsets of all possible bets are jointly
resolvable.  Define a \emph{betting context} to be a set of events
such that bets on all of them are jointly resolvable and to which no
other event can be added without violating this condition.  It is safe
to assume that each betting context is a Boolean algebra, since, if we
can find out whether $E$ occurred at the same time as finding out
whether $F$ occurred, then we can also determine whether they both
occurred, whether either one of them occurred, and whether they failed
to occur, so we can define the usual logical notions of AND, OR and
NOT.  However, unlike in conventional probability theory, there need
not be a common algebra on which all of the events that occur in
different betting contexts are jointly defined.  Because of this, the
Dutch book argument has normative force within a betting context, but
it does not tell us how probabilities should be related across
different contexts.  Therefore, our degrees of belief should be
represented by a set of probability distributions $p(E|\mathcal{B})$,
one for each betting context $\mathcal{B}$\footnote{Despite the
  notation, $p(E|\mathcal{B})$ is not a conditional probability
  distribution because there need not be a common algebra on which all
  the events are defined.  Some authors do not consider this to be a
  generalization of probability theory \cite{Greaves2010,
    Wallace2012}, since all we are saying is that we have a bunch of
  probability distributions rather than just one.  However, such
  systems can display nonclassical features such as violations of Bell
  inequalities and no-cloning \cite{Barrett2007} so they are worthy of
  the name ``generalization'' if anything is.}.

This framework can be applied to quantum theory where the betting
contexts represent sets of measurements that can be performed together
at the same time.  The details of this are rather technical, so they
are relegated to appendix \ref{gleason}.  The probabilities that
result from this are more general than those allowed by quantum
theory.  To get uniquely down to the Born rule, we need an extra
constraint, known as \emph{noncontextuality}.  This says that there
are certain pairs of events from different betting contexts, $E \in
\mathcal{B}$ and $F \in \mathcal{B}'$, that must always be assigned
the same probability $p(E|\mathcal{B}) = p(F|\mathcal{B}')$.
Therefore, we need to explain how such additional constraints can be
understood.

\section{Noncontextuality in subjective Bayesianism}

One option is that noncontextuality could simply be posited as an
additional fundamental principle.  Previous Dutch book arguments for
the Born rule have done essentially this \cite{Caves2002,
  Pitowsky2003}.  However, subjective Bayesians do not accept
fundamental constraints on probabilities beyond those required by
rationality.  Imposing such constraints would be like saying that you
are allowed to construct a logical argument providing one of your
starting premises is ``the car is red'', but if you start from ``the
car is yellow'' then any argument you make is logically invalid.
Additional constraints on probabilities are contingent facts about
your state of belief, just as logical premises are contingent facts
about the world.  Therefore, noncontextuality needs to be derived in
some way.

One possibility is that noncontextuality follows from logical
equivalence, i.e. if quantum theory always assigns the same
probability to $E$ in context $\mathcal{B}$ and $F$ in $\mathcal{B}'$
then these should be regarded as equivalent logical statements, in the
same sense that $E$ and $\text{NOT} (\text{NOT} E)$ are equivalent in
a Boolean algebra\footnote{Pitowsky attempts to argue along these
  lines \cite{Pitowsky2003}, unsuccessfully in my view.}.  Logical
equivalence implies that it ought to be possible to construct a Dutch
book that results in a sure loss if $p(E|\mathcal{B}) \neq
p(F|\mathcal{B}')$.  This can only be done if you are willing to
accept that the occurrence of $E$ in betting context $\mathcal{B}$
makes it necessary that $F$ would have occurred had the betting
context been $\mathcal{B}'$ and vice versa.  If this is the case, then
you will agree that a bet made on $F$ in betting context
$\mathcal{B}'$ should also pay out if the betting context was in fact
$\mathcal{B}$ and the event $E$ occurred and vice versa.  If this is
the case, then the bookie can construct a Dutch book against
$p(E|\mathcal{B}) < p(F|\mathcal{B}')$ by buying a ticket from you
that pays out on $E$ and selling a ticket that pays out on $F$.  The
payouts on these tickets will be the same, so you will lose money in
this transaction.  By exchanging the roles of $E$ and $F$, there would
be a Dutch book against $p(E|\mathcal{B}) > p(F|\mathcal{B}')$ as
well.

This strategy hinges on whether it is reasonable to make
counterfactual assertions, i.e. assertions about what would have
happened had the betting context been different.  However, ``it from
bit'' declares such counterfactuals meaningless because it says that
there is no answer to questions that have not been asked.  Even if we
do not accept ``it from bit'', the Kochen-Specker theorem
\cite{Kochen1967} implies that counterfactual assertions cannot all
respect noncontextuality, i.e. there would have to be pairs of events
$E\in \mathcal{B}$ and $F\in\mathcal{B}'$ such that if $E$ occurs in
$\mathcal{B}$ then $F$ would not have occurred in $\mathcal{B}'$ even
though quantum theory asserts that $p(E|\mathcal{B}) =
p(F|\mathcal{B}')$ always holds.  We conclude that noncontextuality of
probability assignments cannot be a rationality requirement.

Another possibility is that noncontextuality could be adopted simply
because we have performed many quantum experiments and have always
observed relative frequencies in accord with the Born rule.  Although
probabilities are not identified with relative frequencies in
subjective Bayesianism, it still offers an account of statistical
inference wherein observing relative frequencies causes probabilities
to be updated.  If certain technical conditions hold, probability
assignments will converge to the observed relative frequency in the
limit of a large number of trials.  Therefore, we could assert that
noncontextuality is a brute empirical fact\footnote{This has been
  suggested in the context of the many-worlds interpretation
  \cite{Greaves2010}.}.

The problem with this is that it provides no explanation of why
noncontextuality holds.  If we accept this, we might as well just give
up and say that the only reason why we believe any physical theory is
because it matches the observed relative frequencies.  It would be
like saying that the reason why the Maxwell-Boltzmann distribution
applies to a box of gas is because we have sampled many molecules from
such boxes and always found them to be approximately Maxwell-Boltzmann
distributed.  This belies the important explanatory role of stationary
distributions in equilibrium statistical mechanics, and would be of no
help in understanding why nonequilibrium systems tend to equilibrium.
Similarly, the Born rule appears to be playing an important structural
role in quantum theory that calls for an explanation.

The remaining option is to view noncontextuality as arising from
physical, as opposed to logical, equivalence.  The Dutch book
rationality criterion is usually expressed as the requirement that you
should not enter into bets that lead to a sure loss by logical
necessity, but it is equally irrational to enter into a bet that you
believe will lead to a sure loss, whether or not that belief is a
logical necessity.  Because of this, the argument that you should
assign probability one to the certain event equally applies to events
that you only believe to be certain, regardless of whether that belief
is correct.  Now, belief in the laws of physics entails certainty
about statements that follow from the laws so this can be the origin
of constraints on probability assignments.

To illustrate, suppose you believe that Newtonian mechanics is true
and that there is a single particle system with a given Hamiltonian.
This means that you are committed to propositions of the form ``If the
particle initially occupies phase space point $(x_0,p_0)$ then at time
$t$ it occupies the solution to Hamilton's equations $(x(t),p(t)) $
with initial condition $(x_0,p_0)$''.  If you bet on such propositions
at anything more or less than even odds then you believe that you will
lose money with certainty.  Importantly, this type of argument can
also imply constraints on events that you are not certain about.  For
example, if you assign a phase space region some probability and then
compute the endpoints of the trajectories for all points in that
region at a later time then the region formed by the end points must
be assigned the same probability at that later time.  This shows that
the need to assign equal probabilities to different events can
sometimes be derived from the laws of physics.

Crucially, this sort of argument can only really be made to work if
there is an objectively existing external reality.  There needs to be
some sort of ``quantum stuff'' such that events that are always
assigned the same probability correspond to physically equivalent
states of this stuff.  In the context of the many-worlds
interpretation, the Deutsch-Wallace \cite{Deutsch1999, Wallace2007,
  Wallace2010, Wallace2012} and Zurek \cite{Zurek2003, Zurek2005}
derivations of the Born rule are arguments of this type, where the
quantum stuff is simply the wavefunction.

\section{``It from bit'' or ``bit from it''?}

We have arrived at the conclusion that noncontextuality must be
derived in terms of an analysis of the things that objectively exist.
This implies a realist view of physics, or in other words ``bit from
it'', which seems to conflict with ``it from bit''.  Fortunately, this
conflict is only apparent because ``it'' is being used in different
senses in ``it from bit'' and ``bit from it''.  The things that
Wheeler classifies as ``it'' are things like particles, fields and
spacetime.  They are things that appear in the fundamental ontology of
classical physics and hence are things that only appear to be real
from our perspective as classical agents.  He does not mention things
like wavefunctions, subquantum particles, or anything of that sort.
Thus, there remains the possibility that reality is made of quantum
stuff and that the interaction of this stuff with our question asking
apparatus, also made of quantum stuff, is what causes the answers
(particles, fields, etc.) to come into being.  ``It from bit'' can be
maintained in this picture provided the answers depend not only on the
state of the system being measured, but also on the state of the stuff
that comprises the measuring apparatus.  Thus, we would end up with
``it from bit from it'', where the first ``it'' refers to classical
ontology and the second refers to quantum stuff.

\section{Conclusion}

On the subjective Bayesian view, ``it from bit'' implies that
probability theory needs to be generalized, which is in accord with
the observation that quantum theory is a generalized probability
theory.  However, ``it from bit'' does not get us all the way to the
quantum probability rule.  A subjective Bayesian analysis of
noncontextuality indicates that it can only be derived within a
realist approach to physics.  At present, this type of derivation has
only been carried out in the many-worlds interpretation, but I expect
it can be made to work in other realist approaches to quantum theory,
including those yet to be discovered.

\section{Addendum}

In editing this essay for publication, I wanted to hew as closely as
possible to the version submitted to the contest, so I have decided to
address the discussion that occurred on the FQXi website in this
addendum.  I also address some comments made by Kathryn Laskey in
private correspondence, because I think she addressed one of the
issues particularly eloquently.  I am grateful to my colleagues and
co-entrants for their thoughtful comments.  It would be impossible to
address all of them here, so I restrict attention to some of the most
important and frequently raised issues.  Further details can be found
on the FQXi comment thread \cite{Comment2013}.

\subsection{Noncontextuality}

Both Jochen Szangolies and Ian Durham expressed confusion at my usage
of the term ``noncontextuality'', which derives from Gleason's theorem
\cite{Gleason1957}.  Due to the Kochen-Specker theorem
\cite{Kochen1967}, it is often said that quantum theory is
``contextual'', so how can this be reconciled with my claim that the
Born rule is ``noncontextual''?

In Gleason's theorem, noncontextuality means that the same probability
should be assigned to the same projection operator, regardless of the
context that it is measured in.  Here, by context, I mean the other
projection operators that are measured simultaneously.  So, as in the
example given in the technical endnotes, $\ket{2}$ should receive the
same probability regardless of whether it is measured as part of the
basis $\{\ket{0}, \ket{1}, \ket{2} \}$ or the basis $\{\ket{+},
\ket{-}, \ket{2}\}$, where $\ket{\pm} = \frac{1}{\sqrt{2}} \left (
  \ket{0} \pm \ket{1} \right )$.  From the perspective of this essay,
Gleason's theorem says that, for Hilbert spaces of dimension three or
larger, the only probability assignments compatible with both the
Dutch book constraints and noncontextuality are those given by the
Born rule.  In this sense the Born rule is ``noncontextual'' and
indeed it is the only viable probability rule that is.

On the other hand, the Kochen-Specker theorem concerns the assignment
of definite values to the outcomes of measurements.  Instead of
assigning probabilities to projectors, the aim is to assign them
values $0$ or $1$ in such a way that, for any set of orthogonal
projectors that can occur together in a measurement, exactly one of
them gets the value $1$.  This is to be done noncontextually, which
means that whatever value a projector is assigned in one measurement
context, it must be assigned the same value in all other contexts in
which it occurs.  The Kochen-Specker theorem says that this cannot be
done.

The two theorems are related because $0$ and $1$ are examples of
probability assignments, albeit extremal ones.  As first pointed out
by Bell \cite{Bell1966}, Gleason's theorem actually implies the
conclusion of the Kochen-Specker theorem by the following argument.
For any quantum state, the Born rule never assigns $0/1$ probabilities
to every single projector.  Gleason's theorem implies that, in
dimension three and higher, the only noncontextual probability
assignments are given by the Born rule.  Therefore, for these
dimensions, there can be no noncontextual probability assignment that
only assigns $0/1$ probabilities.

From this it should be apparent that the noncontextuality assumption
of the Kochen-Specker theorem is the same as in Gleason's theorem,
only that it is specialized to $0/1$ probability assignments.  The
additional assumption that the probabilities must be either $0$ or $1$
is called \emph{outcome determinism}, so the Kochen-Specker theorem
shows that it is impossible to satisfy both outcome determinism and
noncontextuality at the same time (in addition to the Dutch book
constraints).

Based on this, people often loosely say that the Kochen-Specker
theorem shows that quantum theory is ``contextual'' and this is the
source of the confusion.  However, adopting contextual value
assignments is only one way of resolving the contradiction entailed by
the Kochen-Specker theorem, the other being to drop outcome
determinism.  It is therefore perfectly consistent to say that the
Born rule is noncontextual but that any model that assigns definite
values to every observable cannot be.

\subsection{Scientific realism}

Scientific realism is the view that our best scientific theories
should be thought of as describing an objective reality that exists
independently of us.  My argument ends up endorsing the realist
position, as it concludes that the world must be made of some
objectively existing ``quantum stuff''.

There are good a priori reasons for believing in scientific realism
that are independent of the specifics of quantum theory, and hence
independent of the argument given in this essay (see
\cite{Chakravartty2014} for a summary and \cite{Ladyman2002} for a
more detailed treatment of these arguments).  Most people who believe
in scientific realism are probably swayed by these arguments rather
than anything to do with the details of quantum theory.

As pointed out by Ken Wharton, this would seem to open the possibility
of short-circuiting my argument.  Why not simply make the case for
scientific realism via one or more of the a priori arguments?  From
this it follows that the world must consist of some objectively
existing stuff, and hence ``bit from it''.

Whilst I agree that this is a valid line of argument, my intention was
not to provide an argument that would convince realists, for whom
``bit from it'' is a truism.  It is evident from the popularity of
interpretations of quantum theory that draw inspiration from the
Copenhagen interpretation, which I collectively call \emph{neo
  Copenhagen} interpretations, that not everyone shares such strong
realist convictions.  Wheeler's ``it from bit'' is usually read as a
neo-Copenhagen principle.  It says that what we usually call reality
derives from the act of making measurements rather than from something
that exists independently of us.  As I argue in the essay, ``it from
bit'' can be given a more realist spin by interpreting ``it'' as
referring to an emergent, effective classical reality rather than to
the stuff that the world is made of at the fundamental level.
Nevertheless, most endorsers of ``it from bit'' are likely to have the
neo Copenhagen take on it in mind.

The most effective way of arguing against any opponent is to start
from their own premises and show that they lead to the position they
intend to oppose.  This is much more effective than arguing against
their premises on a priori grounds as it is evident from the fact that
they have chosen those premises that the opponent does not find such a
priori arguments compelling.  My aim here is to do this with ``it from
bit''---a premise accepted by many neo Copenhagenists---and to argue
that it needs to be supplemented with realism, or ``bit from it'', in
order to obtain a compelling derivation of the Born rule.  This
presents a greater challenge to the neo Copenhagen view than simply
rehashing the existing arguments for realism.  I expect my fellow
realists to find this line of argument overly convoluted, but it is
not really aimed at them.

\subsection{Has probability theory really been generalized?}

In this essay, I argued that ``it from bit'' requires a generalization
of probability theory.  Specifically, I argued that there are a number
of different betting contexts $\mathcal{B}_1, \mathcal{B}_2, \ldots$,
that within each betting context the Dutch book argument implies a
well defined probability measure over the Boolean algebra of events in
that context, but that it does not imply any constraints on events
across different betting contexts.  This gives rise to a theory in
which there are a number of different Boolean algebras, each of which
has its own probability measure, instead of there being just one
probability measure over a single Boolean algebra.  Giacomo Mauro
D'Ariano, Howard Barnum and Kathryn Laskey (the latter in private
correspondence) questioned whether it is really necessary to think of
this as a generalization of probability theory.

D'Ariano's method for preserving probability theory is to assign
probabilities to the betting contexts themselves.  That is, we can
build a sample space of the form $(\mathcal{B}_1 \times
\Omega_{\mathcal{B}_1}, \mathcal{B}_2 \times \Omega_{\mathcal{B}_2},
\ldots)$, where $\Omega_{\mathcal{B}_j}$ is the sample space
associated with betting context $\mathcal{B}_j$.  We can then just
specify an ordinary probability measure over this larger space, and
the separate probability measures for each context would then be
obtained by conditioning on $\mathcal{B}_j$.

I admit that this can always be done formally, but conceptually one
might not want to regard betting contexts as the kind of thing that
should be assigned probabilities.  They are defined by the sequences
of questions that we decide to ask, so one might want to regard them
as a matter of ``free choice''.  To avoid the thorny issue of free
will, we can alternatively imagine that the betting context is
determined by an adversary.  Recall that, for a subjective Bayesian,
assigning a probability to an event means being willing to bet on that
event at certain odds.  Therefore, assigning probabilities to betting
contexts means you should be willing to bet on which context will
occur.  However, if the bookie is also the person who gets to choose
the betting context after all such bets are laid, then she can always
do so in such a way as to make your probability assignments to the
betting contexts as inaccurate as possible.  Therefore, there are at
least some circumstances under which it would not be meaningful to
assign probabilities to betting contexts.

Laskey's response is quite different.  She simply denies that what I
have described deserves the name ``generalization of probability
theory''.  Since her comments were made in private communication, with
her permission I reproduce them here.

\begin{quotation}
  Let me first take issue with your statement that quantum theory
  requires generalizing probability theory because the Boolean
  algebras of outcomes are different in different betting contexts.
  Dependence of the Boolean algebra of outcomes on the betting context
  is by no means restricted to quantum theory. It happens all the time
  in classical contexts -- in fact, it's a fixture of our daily life.
  Ever see the movie ``Sliding Doors''?  The Boolean algebra of outcomes
  I face today would be totally different had I not chosen to marry my
  husband; had I taken a different job when I came out of grad school;
  had my husband and I not had four children; had I not chosen an
  academic career; had I not put myself into a position in which other
  people depend on me to put food on the table; or any number of other
  might-have-beens in my life.

  Consider, for example, a town facing the question of whether to zone
  a given area for residential development or to put a wind farm
  there.  If the town chooses residential development, we might have,
  for example, a probability distribution over a Boolean algebra of
  values of the average square footage of homes in the area. There
  would be no such Boolean algebra if we build the wind farm. (I am
  specifically considering averages because they are undefined when N
  is zero.)  If we choose the wind farm, we would have a distribution
  over the average daily number of kilowatt hours of wind-powered
  electricity generated by the wind farm. There would be no such
  Boolean algebra if we choose the residential development.  What is
  the intrinsic difference between this situation and the case of a
  quantum measurement, in which the algebra of post-measurement states
  depends on the experiment the scientist chooses to conduct?

  Just about any time we make a decision, the Boolean algebra of
  possible future states of the world is different for each choice we
  might make.  Decision theorists are accustomed to this dependence of
  possible outcomes on the decision. It does not mean we need to
  generalize probability theory. It simply means we have a different
  Boolean algebra conditional on some contexts than conditional on
  others.
\end{quotation}

In some ways this is a matter of semantics.  I argue in the essay that
the breakdown of some of the usual conventions of probability theory
is commonplace and should not be surprising.  We just disagree on
whether this deserves the name ``generalization''.

My argument for a generalization of probability theory is mainly
directed against dogmatic Bayesians who endorse the view that ordinary
probability theory on a single Boolean algebra is not to be violated
on pain of irrationality.  There are plenty of dogmatic Bayesians
still around.  If modern Bayesians have a more relaxed attitude then
that is all to the good as far as I am concerned.  However, I do think
it is worth making the argument specifically in the context of
physics, as physicists are often a bit timid about drawing
implications for the foundations of probability from their subject,
and I do not think they should be if violations of the standard
framework are commonplace.

I therefore do not wish to spill too much ink over whether or not the
bare-bones theory of multiple Boolean algebras should be called a
generalization of probability theory.  However, quantum theory has
much more structure than this, in the form of Hilbert space structure
and the noncontextuality requirement.  For me, the more important
question is whether quantum theory should be viewed as a
generalization of probability theory.

\subsection{Must quantum theory be viewed as a generalization of
  probability theory?}

The short answer to this is no.  The underdetermination of theory by
evidence implies that there will always be several ways of formulating
a theory that are empirically equivalent.  We can always apply
D'Ariano's trick or take Laskey's view, since they apply to any set of
probabilities on separate Boolean algebras, and quantum theory is just
a restriction on that set.  Therefore, I cannot argue that it is a
logical necessity to view quantum theory as a generalized probability
theory, but I can argue that it is more elegant, simpler, productive,
etc. to do so.

As an analogy, note that it is also not logically necessary to view
special relativity as ruling the existence of a luminiferous ether.
Instead, one can posit that there exists an ether, that it picks out a
preferred frame of reference in which it is stationary, but that
forces act upon objects in such a way to make it impossible to detect
motion relative to the ether, e.g. they cause bodies moving relative
to the ether to contract in just such a way as to mimic relativistic
length contraction.  This theory makes the exact same predictions as
special relativity and is often called the Lorentz ether
theory\footnote{It is similar to the theory in which Lorentz first
  derived his eponymous transformations, although, unlike the theory
  described here, the actual theory proposed by Lorentz failed to
  agree with special relativity in full detail.}.  Special relativity
is normally regarded as superior to the Lorentz ether theory because
the latter seems to require a weird conspiracy of forces in order to
protect the existence of an entity that cannot be observed.  The
former has proved to be a much better guide to the future development
of physics.  What I want to argue is that not adopting a view in which
quantum theory is a generalization of probability theory is analogous
to adopting the Lorentz ether theory, i.e. it is consistent but a poor
guide to the future progress of physics.

When we add Hilbert spaces and noncontextuality into the mix,
Gleason's theorem implies that our beliefs can be represented by a
density operator $\rho$ on Hilbert space, at least if the Hilbert
space is of dimension three of higher.  I have argued elsewhere that
regarding the density operator as a true generalization of a classical
probability distribution leads to an elegant theory which unifies a
lot of otherwise disparate quantum phenomena \cite{Leifer2013}.  Here,
I will confine myself to a different argument, based on the quantum
notion of entropy.

Classically, the entropy of a probability distribution $\bm{p} =
(p_1,p_2,\ldots,p_n)$ over a finite space is given by 
\begin{equation}
  H(\bm{p})-\sum_j p_j \log p_j.
\end{equation}  
Up to a multiplicative constant, this describes both the Shannon
(information theoretic) entropy and the Gibbs (thermodynamic) entropy.
In other words, it describes the degree of compressibility of a string
of digits drawn from independent instances of the probability
distribution $\bm{p}$ and also quantifies the amount of heat that must
be dissipated in a thermodynamic transformation.  In quantum theory,
the entropy of a density operator is given by the von Neumann entropy,
\begin{equation}
  S(\rho) = -\Tr{\rho \log \rho},
\end{equation} 
which is the natural way of generalizing the classical entropy if you
think of density operators as the quantum generalization of
probability distributions.  It turns out that this plays the same role
in quantum theory as the classical entropy does in classical theories,
i.e. it is both the information theoretic entropy, quantifying the
compressibility of quantum states drawn from a source described by
$\rho$, and it is the thermodynamic entropy, quantifying the heat
dissipation in a thermodynamic transformation.  

Now, this definition of quantum entropy only really makes sense on the
view that density operators are generalized probability measures.
What would we get if we took D'Ariano or Laskey's views instead?  

On D'Ariano's view we have a well-defined classical probability
distribution, just over a larger space that includes the betting
contexts.  If this is just an ordinary classical probability
distribution then arguably we should just use the formula for the
classical entropy, although one might want to marginalize over the
betting contexts first.  This is not the von Neumann entropy, and it
does not seem to quantify anything of relevance to quantum information
or thermodynamics.  

On Laskey's view we just have a bunch of unrelated probability
distributions over different betting contexts.  Should we take the
entropy of just one of these and, if so, which one?  None of them
seems particularly preferred.  Should we take some kind of weighted
average of all of them and, if so, what motivates the weighting given
that betting contexts are not assigned probabilities?  Arguably the
only relevant betting context is the one we actually end up in, so one
should just apply the classical entropy formula to this context, but
this is unlikely to match the von Neumann entropy\footnote{It will
  match if we are lucky enough to choose the context that minimizes
  the classical entropy, but again there is no motivation for doing
  this in Laskey's approach.}.

Now admittedly, there is probably some convoluted way of getting to
the von Neumann entropy in these other approaches, just as there is a
way of understanding the Lorentz transformations in the Lorentz ether
theory, but I expect that it would look ad hoc compared to treating
density operators as generalized probability distributions.

In summary, I am arguing that quantum theory should be regarded as a
bona fide generalization of probability theory, not out of logical
necessity, but because doing so gives the right quantum
generalizations of classical concepts.  People who view things in this
way are liable to make more progress in quantum information theory,
quantum thermodynamics, and beyond, than those who do not.  In this
sense I think the situation is analogous to adopting special
relativity over Lorentz ether theory.

\bibliographystyle{Science}
\bibliography{FQXI2013}

\appendix

\section{Technical Endnotes}

\label{gleason}

In general, a betting context $\mathcal{B}$ is a Boolean algebra,
which we take to be finite for simplicity.  All such algebras are
isomorphic to the algebra generated by the subsets of some finite set
$\Omega_{\mathcal{B}}$, where AND is represented by set intersection,
OR by union, and NOT by complement.

In quantum theory, a betting context corresponds to a set of
measurements that can be performed together that is as large as
possible.  A measurement is represented by a self-adjoint operator $M$
and all such operators have a spectral decomposition of the form
\begin{equation}
  M = \sum_j \lambda_j \Pi_j,
\end{equation}
with eigenvalues $\lambda_j$ and orthogonal projection operators
$\Pi_j$ that sum to the identity $\sum_j \Pi_j = I$.  The eigenvalues
are the possible measurement outcomes and, when the system is assigned
the density operator $\rho$, the Born rule states that the outcome
$\lambda_j$ is obtained with probability
\begin{equation}
  p(\lambda_j) = \Tr{\Pi_j \rho}.
\end{equation}

The eigenvalues just represent an arbitrary labelling of the
measurement outcomes, so a measurement can alternatively be
represented by a set of orthogonal projection operators $\{\Pi_j\}$
that sum to the identity $\sum_j \Pi_j = I$, which is sometimes known
as a \emph{Projection Valued Measure (PVM)}\footnote{More generally,
  we could work with Positive Operator Valued Measures (POVMs) or sets
  of consistent histories, but this would not substantially change the
  arguments of this essay.}.

Two PVMs $A = \{\Pi_j\}$ and $B = \{\Pi'_j\}$ can be measured together
if and only if each of the projectors commute, i.e. $\Pi_j \Pi'_k =
\Pi'_k \Pi_j$ for all $j$ and $k$.  If this is the case then
$\Pi_j\Pi'_k$ is also a projector and $\sum_{jk}\Pi_j\Pi_k = I$.
Therefore, one way of performing the joint measurement is to measure
the PVM $C = \{\Pi''_{jk}\}$ with projectors $\Pi''_{jk} =
\Pi_j\Pi'_k$ and, upon obtaining the outcome $(jk)$, report the
outcome $j$ for $A$ and $k$ for $B$.  This fine graining procedure can
be iterated by adding further commuting PVMs and forming the product
of their elements with those of $C$.  The procedure terminates when
the resulting PVM is as fine grained as possible and this will happen
when it consists of rank-$1$ projectors onto the elements of an
orthonormal basis.  The outcome of any other commuting PVM is
determined by coarse graining the projectors onto the orthonormal
basis elements.

Therefore, in quantum theory, we can take the sets
$\Omega_{\mathcal{B}}$ that generate the betting contexts
$\mathcal{B}$ to consist of the elements of orthonormal bases.  An
event $E \in \mathcal{B}$ is then a subset of the basis elements and
corresponds to a projection operator $\Pi_E = \sum_{\ket{\psi} \in E}
\proj{\psi}$.  The Boolean operations on $\mathcal{B}$ can be
represented in terms of these projectors as
\begin{itemize}
\item Conjunction: $G = E \, \text{AND} \, F \,\, \Rightarrow \,\,
  \Pi_G = \Pi_E\Pi_F$
\item Disjunction: $G = E \, \text{OR} \, F \,\, \Rightarrow \,\,
  \Pi_G = \Pi_G + \Pi_F - \Pi_G\Pi_F$, which reduces to $\Pi_G = \Pi_G
  + \Pi_F$ when $E \cap F = \emptyset$.
\item Negation: $G = \text{NOT} \, E \,\, \Rightarrow \,\, \Pi_G = I -
  \Pi_E$.
\end{itemize}

From the Dutch book argument applied within a betting context, we have
that our degrees of belief should be represented by a set of
probability measures $p(E|\mathcal{B})$ satisfying
\begin{itemize}
  \item For any event $E \subseteq \Omega_{\mathcal{B}}$,
    $p(E|\mathcal{B}) \geq 0$.
  \item For the certain events $\Omega_{\mathcal{B}}$,
    $p(\Omega_{\mathcal{B}}|\mathcal{B}) = 1$
  \item For disjoint events within the same betting context $E,F
    \subseteq \Omega_{\mathcal{B}}$, $E \cap F = \emptyset$, $p(E \cup
    F|\mathcal{B}) = p(E|\mathcal{B}) + p(F|\mathcal{B})$.
\end{itemize}
The Born rule is an example of such an assignment, and in this
language it takes the form
\begin{equation}
  p(E|\mathcal{B}) = \Tr{\Pi_E \rho}.
\end{equation}

The Born rule also has the property that the probability only depends
on the projector associated with an event, and not on the betting
context that it occurs in.  For example, in a three dimensional
Hilbert space, consider the betting contexts $\Omega_{\mathcal{B}} =
\{\ket{0},\ket{1},\ket{2}\}$ and $\Omega_{\mathcal{B}'} =
\{\ket{+},\ket{-},\ket{2}\}$, where $\ket{\pm} = \frac{1}{\sqrt{2}}
\left ( \ket{0} \pm \ket{1} \right )$.  The Born rule implies that
$p(\{\ket{2}\}|\mathcal{B}) = p(\{\ket{2}\}|\mathcal{B}')$ and also
that $p(\{\ket{0},\ket{1}\}|\mathcal{B}) =
p(\{\ket{+},\ket{-}\}|\mathcal{B}')$ because, in each case, the events
correspond to the same projectors.  The Dutch book argument alone does
not imply this because it does not impose any constraints across
different betting contexts.

A probability assignment is called \emph{noncontextual} if
$p(E|\mathcal{B}) = p(F|\mathcal{B}')$ whenever $\Pi_E = \Pi_F$.
Gleason's theorem \cite{Gleason1957} says that, in Hilbert spaces of
dimension $3$ or larger, noncontextual probability assignments are
exactly those for which there exists a density operator $\rho$ such
that $p(E|\mathcal{B}) = \Tr{\Pi_E \rho}$, i.e. they must take the
form of the Born rule.  Therefore, the Born rule follows from the
conjunction of the Dutch book constraints and noncontextuality, at
least in Hilbert spaces of dimension $3$ or greater.

\end{document}